\documentclass[twoside]{article}

\usepackage{lipsum} 

\usepackage[round]{natbib}
\usepackage[sc]{mathpazo} 
\usepackage[T1]{fontenc} 
\usepackage{microtype} 

\usepackage[hmarginratio=1:1,top=32mm,columnsep=20pt]{geometry} 
\usepackage{multicol} 
\usepackage[hang, small,labelfont=bf,up,textfont=it,up]{caption} 
\usepackage{booktabs} 
\usepackage{float} 

\usepackage[hidelinks]{hyperref}

\usepackage{lettrine} 
\usepackage{paralist} 

\usepackage{abstract} 

\usepackage{titlesec} 
\renewcommand\thesection{\Roman{section}} 
\renewcommand\thesubsection{\Roman{subsection}} 
\titleformat{\section}[block]{\large\scshape\centering}{\thesection.}{1em}{} 
\titleformat{\subsection}[block]{\large}{\thesubsection.}{1em}{} 

\usepackage{fancyhdr} 
\usepackage{graphicx}
\DeclareGraphicsExtensions{.jpeg}

\title{\vspace{-15mm}\fontsize{21pt}{10pt}\selectfont\textbf{Characterization of the Community Structure of Large-Scale Functional Brain Networks  During Ketamine-Medetomidine  Anesthetic Induction  }} 

\author{
\large
\textsc{Eduardo C. Padovani}
\thanks{Email: \texttt{eduardo.padovani@alumni.usp.br}}\\[2mm] 
\normalsize 
\vspace{-5mm}
}
\date{}

\begin{document}

\maketitle 

\thispagestyle{fancy} 


\begin{abstract}

\noindent \ One of the main goals of neuroscience is to understand how an organism's cognitive capacities or physiological states are potentially related to brain activities that involve the interaction of multiple brain structures and cortical areas. A key feature of functional brain networks is that they are modularly structured, and this modular architecture is regarded as accounting for a range of properties and functional dynamics. In the neurobiological context, communities may indicate brain regions involved in the same activity, representing neural-segregated processes. Several studies have demonstrated the modular character of the organization of brain activities. However, empirical evidence regarding its dynamics and relation to different levels of consciousness has not yet been reported. In this context, this research sought to characterize the community structure of functional brain networks during an anesthetic induction process. The experiment was based on intracranial recordings of the neural activities of an old-world macaque of the species Macaca fuscata during a Ketamine-Medetomidine anesthetic induction process. Changes were observed within approximately one and a half minutes after administering the anesthetics, revealing a transition in the community structure. The awake state was characterized by large clusters involving the frontal and parietal regions. In contrast, the anesthetized state was marked by the presence of communities in the primary visual and motor cortices, while the areas of the secondary associative cortex were the most affected. The results report the influence of general anesthesia on the structure of functional clusters, contributing to understanding some novel aspects of the neural correlates of consciousness.

\end{abstract}


\begin{multicols}{2} 

\section{Introduction}
\lettrine[nindent=0em, lines=3]{M}odern network science is beginning to influence our ideas on understanding the processes that underlie brain activities \citep{bullmore2009complex, sporns2014contributions}. Independent studies carried out with heterogeneous methodological approaches \citep{van2008small,power2013evidence,
padovani2016characterization}, with an encouraging degree of confidence, indicate that functional brain networks exhibit specific topological features, which supports the consensus that a certain functional organization architecture is necessary for higher brain functions to arise \citep{stam2012organization}. Complex network tools \citep{rubinov2010complex} provide ways to characterize and quantify generic organizational principles in the nervous system. In recent years, this approach has been extensively used in neuroscience to understand brain neural network organization.
A key organizational feature identified in the brain's functional and structural networks is the modular structure \citep{meunier2010modular}, specified by the presence of modules, also called communities \citep{newman2006modularity}. \textit{Communities} are defined as subsets of nodes that are highly connected among themselves and less densely connected to other nodes of the network
 (see \hyperlink{FIGURE1}{$Figure \cdot 1$} for a schematic representation of the concept of communities in a network).

\begingroup
    \centering
    \includegraphics[width=8 cm]{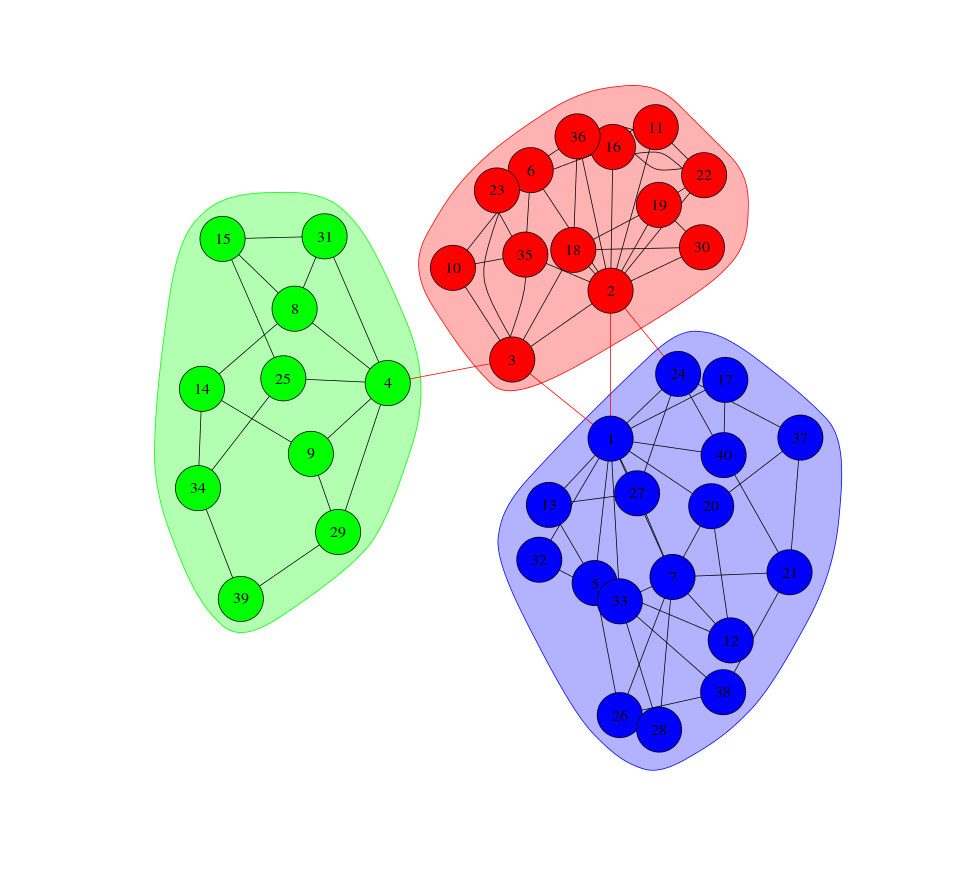}
    \captionof{figure}{A schematic representation of a generic network with 40 nodes, possessing three groups of vertexes densely connected among themselves and less densely connected to other vertexes of the network. The community structure (each community represented by a different color) was identified by the algorithm \textit{Walktrap Community
 \citep{pons2006computing}.}  }\label{fig:a}
 \hypertarget{FIGURE1}{}
\endgroup

\vspace*{5mm}

The structure of large-scale functional brain networks is regarded as reflecting the specific way information processing and integration are performed between several brain regions \citep{stam2007graph,sporns2011networks}. It is also believed that those networks constitute a substrate that underlies different types of physical dynamics and emergent properties \citep{bassett2006small}. This way, the modular architecture is considered beneficial for the design of the nervous system once mechanisms supported by modular systems are believed to play a role in the brain \citep{meunier2010modular}. Modular networks display small-world properties; a high density of connections in the same module favors locally segregated processing, while the connections between communities lead to the property of short path length that may support globally integrated processing \citep{sporns2004organization,meunier2010modular}. Modular networks aim to produce a time-scale separation, accounted for by fast intra-modular and slow inter-modular processes \citep{pan2009modularity}. The presence of communities is thought to allow neural activities to prevail locally encapsulated and, at the same time, to maintain a dynamical balance between the extremes of rapidly dying out or spreading throughout the whole network \citep{kaiser2007criticality,kaiser2010optimal}.\\

The modular character highlights a particular aspect of the brain's functional connectivity organization, revealing its tendency to form segregated subsystems with specialized functional properties \citep{sporns2004organization,meunier2010modular,sporns2011networks}. Such functional clusters may also be seen as manifestations of integrated and differentiated neural processes involving distinct brain areas, which may be potentially related to conscious experiences or cognitive capacities \citep{edelman2001consciousness,tononi1998consciousness,
 edelman2013consciousness}. There are also theoretical predictions claiming that distinct conscious levels or cognitive demands may be associated with alterations in the modular character of brain activities \citep{dehaene1998neuronal}. 


\subsubsection*{General Anesthesia}

Understanding general anesthesia is highly relevant for both medicine and neuroscience.
In clinical medicine, anesthetics are among the most commonly used neurotropic drugs in the world \citep{uhrig2014cerebral}. Every day, thousands of people undergo procedures that involve the use of general anesthesia \citep{schwartz2010general}.
For neuroscience, anesthetic agents constitute experimental tools able to induce different levels of consciousness in a stable and reproductive manner \citep{uhrig2014cerebral}. Thus, they offer an excellent opportunity to study consciousness and the neural correlates of consciousness, providing possibilities to comprehend fundamental processes and phenomena that happen in the brain \citep{hameroff1998toward}.
Given their importance and the fact that very little is known about the neurophysiological mechanisms that underlie this state involving sedation and loss of consciousness \citep{schwartz2010general,lewis2012rapid}, the \textit{Science Magazine}  has pointed out the elucidation of the processes and mechanisms of general anesthesia as one of the 125 most important open questions in science \citep{kennedy2005don}.

Within this context, this study sought to characterize the community structure of functional brain networks estimated serially in short time intervals during an anesthetic induction process, enabling us to infer their structure and dynamics under controlled experimental conditions that involved a drastic and fast reduction in the level of consciousness. This way, the results also reveal empirical evidence regarding aspects of the neural correlates of consciousness.


\section{Methods}

There are hypotheses and predictions that the functional clusters might not be static, and may dynamically change in number, size, and composition involving distinct specialized cortical regions at a time scale typically ranging from hundreds of milliseconds to a few seconds \citep{tononi1998consciousness}.

To verify the spatial location of these functional clusters, as well as their dynamics, a recording technique capable of providing both wide coverage of the cortical surface, and high spatial and temporal resolution is required to resolve the spatial delimitation coverage and follow the dynamics at the time scale at which changes might occur in the brain.

 To address these objectives, we have analyzed a database of cortical electrophysiological activity recordings coming from the \texttt{RIKEN Brain Science Institute}, Saitama, Japan. In this database, neural records were obtained by the \textit{MDR-ECoG} technique \citep{nagasaka2011multidimensional,fukushima2014electrocorticographic}, which is based on a high-density array of ECoG electrodes chronically implanted below the dura mater and positioned over the cortex in a macaque experimental model. The ECoG electrodes continuously covered, with a spatial resolution of 5 mm, the lateral cortical surface of the left cerebral hemisphere and parts of the frontal and occipital medial walls. The cortical electrophysiological activity was recorded at 1KHz of temporal resolution.
 
The database analyzed in the present research is respective to a \textit{Ketamine} and \textit{Medetomidine} anesthetic induction experiment in an animal model subject of the species \textit{Macaca fuscata} \citep{nagasaka2011multidimensional}. This macaque species is one of the main primates used as experimental models in neuroscience and is known to share considerable anatomical and evolutionary similarities with humans \citep{iriki2008neuroscience,isa2009japanese}.


\subsubsection*{Anesthetic Agents}

 Ketamine is a drug able to induce an anesthetic state characterized by the dissociation between the thalamocortical and limbic systems \citep{bergman1999ketamine}. It acts as a non-competitive antagonist of the receptor N-metil-D-asparthate \citep{green2011clinical}. Medetomidine (an agonist of the  alpha-2  adrenergic receptor) was combined with Ketamine to promote muscular relaxation \citep{young1999short}. The antagonist of the Medetomidine, Atipamezole, was used to trigger and promote the recovery process \citep{young1999short}.



\vspace{2.5\baselineskip}

\subsubsection*{Neural Connectivity Estimator}

\vspace{1.5\baselineskip}

As a neural connectivity estimator, a method based on \textit{Granger causality} \citep{granger1969investigating} was used to infer statistical dependencies between the time series of the electrodes. When applied in neuroscience, Granger causality provides an estimate of the exchange of information flow from one cortical area to another \citep{seth2007distinguishing,seth2010matlab}.

\newpage

\subsubsection*{Experimental Procedures - Summary of Steps}

The characterization of the modular structure of functional neural activities performed in this study followed these steps:

\begin{enumerate}

\item A database of cortical electrophysiological activity recorded by the MDR-ECoG technique was used. In the database's experiment, the ECoG electrode's matrix continuously covered an entire brain hemisphere and parts of the medial walls.

\item Each one of the electrodes of the array was considered a vertex of the network and represented the cortical area in which it was positioned.

\item A neural connectivity estimator of Granger causality in the frequency domain was used to estimate values of association between the registers (time series) of the electrodes.

\item An adjacency matrix was assembled, containing all the pairwise association values between the nodes.

\item A community detection algorithm was used to find the community structure of the networks. A criterion was applied to identify highly connected communities to infer the presence of highly integrated and segregated neural processes.

\end{enumerate}

Networks were estimated serially in time intervals of five seconds in five physiological frequency bands along with the experiment. In each frequency band, all the networks in the sequence and their community structure were estimated using the same procedures and parameters. Thus, the alterations observed in the community structure of distinct resolved-in-time networks came from differences in the records of neural activity that occurred during the experiment.

\subsection*{Database - Experimental Procedures }

 The monkey was seated in a proper chair with its head and arms restrained. The neural activity started to be recorded with the monkey awake and with open eyes. Later, the eyes were covered with a patch to prevent a visually evoked response.  After about 10 minutes, a Ketamine-Medetomidine cocktail (5.6mg/Kg of Ketamine + 0.01mg/Kg of Medetomidine) was administered intramuscularly to induce anesthesia. The loss of consciousness point (LOC) was set at the time when the macaque no longer responded to external stimuli (touching the nostrils or opening the hands). After establishing the LOC, neural activity was recorded for about 25-30 minutes. After that period, the antagonist of Medetomidine, Atipamezole (0.05mg/Kg), was administrated intramuscularly to trigger and promote recovery. The monkey recovered from the anesthetic induction, and neural activity was recorded for more some minutes. Heart rate and breathing were monitored throughout the entire experiment.

All data records, experimental, and surgical procedures were performed by researchers from the laboratory of adaptive intelligence at the \texttt{RIKEN - Brain Science Institute}, according to the protocols (No. H24-2-203(4)) approved by the \texttt{RIKEN} ethics committee and the recommendations of the Weatherall report, "The use of non-human primates in research". For further information regarding the experiment, methods, and materials, see \citep{nagasaka2011multidimensional} and (\texttt{http://neurotycho.org}).

\subsection{Signal Processing and Granger Causality in the Frequency Domain}

\subsubsection*{Data Processing}

\begin{enumerate}

\item A reject-band \texttt{IIR-notch} filter was used to attenuate components of the signal at 50Hz.

\item The signal was down-sampled from 1KHz to 200Hz.

\item The signal was divided into windows of 1000 points (equivalent to a five-second recording of neural activity).

\item For each of the 128-time series, the tendency was removed and the average was subtracted.

\item To verify the stationary condition of the time series, the tests \textit{KPSS} \citep{kwiatkowski1992testing} and \textit{ADF} [Augmented Dickey Fuller] \citep{hamilton1989new} were applied.

\end{enumerate}

\subsubsection*{Libraries Used}

For the computation of association values using Granger causality in the frequency domain,  with some adaptations, the following libraries were used: \texttt{MVGC GRANGER TOOLBOX}, developed by Ph.D. Anil Seth (Sussex University, UK), described in \citep{seth2010matlab}, available at \texttt{www.anilseth.com}, and the library \texttt{BSMART toolbox} (\textit{\textbf{B}rain-\textbf{S}ystem for \textbf{M}ultivariate \textbf{A}uto\textbf{R}egressive \textbf{T}imeseries} \texttt{toolbox}) described in \citep{cui2008bsmart} and available at \texttt{www.brain-smart.org}.

\subsubsection*{Computation of Causal Interactions}

\begin{enumerate}

\item Model Order:

To find the model order (number of observations to be used in the regression model), the criteria for the selection of models from Akaike (AIC) and Bayes/Schwartz (BIC) were used.
Both methods returned the order of the model equal to seven.

\item Causal Interactions

At each window of 1000 points, Granger causality in the frequency domain interactions was pair-wise computed among the 128-time series, by the use of the function \texttt{cca\_pwcausal()} (\texttt{MVGC GRANGER TOOLBOX}).

\item Frequency Bands

 Granger causality interactions were calculated in five physiological frequency bands: Delta (0-4Hz), Theta (4-8Hz), Alpha (8-12Hz), Beta (13-30Hz), and Gamma (25-100Hz). 

The interaction values obtained were saved into adjacency matrices.

\end{enumerate}

\subsubsection*{Graphs and Networks}

\begin{enumerate}

\item Assemble Networks

For each sequence of graphs respective to a frequency band, a threshold was chosen, and only the interactions with a magnitude value higher than this threshold were considered edges of the graphs.

\begin{itemize}
\item Delta (0-4Hz), threshold = $0.5$
\item Theta (4-8Hz), threshold = $0.5$
\item Alpha (8-12Hz), threshold = $0.5$
\item Beta (13-30Hz), threshold = $1.5$
\item Gamma (25-100Hz), threshold = $3.75$

\end{itemize}

\end{enumerate}

As discussed in \citep{bullmore2009complex,sporns2011networks}, scientists can use different criteria to determine the threshold parameter. In the present study, due to experimental conditions, each sequence of networks contained graphs with distinct connectivity. Thresholds were chosen in such a way as to prevent graphs with lower connectivity in each sequence from presenting many disconnected parts or vertexes, which might introduce distortions in the analysis.
\vspace*{2mm}

After obtaining non-weighted graphs, the directions of the edges were removed, resulting in undirected and non-weighted networks. Those networks were used for the analysis of the community structure.

\subsection{Community Structure}

Given the community structure of the networks, two different criteria were used in order to distinguish communities that might represent highly integrated and segregated neural processes.\\

\noindent \textbf{Autor's note:} Carefully observing the community structure of several estimated networks plotted through different graph layouts, it was noted that functional brain networks are heterogeneous, presenting in the same graph groups of highly connected vertexes, forming dense agglomerates, and also sets of less densely connected vertexes (sparse parts of the graph). As the author intended to verify the existence, structure, and dynamics of highly integrated and segregated neural processes, a criterion was used to discriminate the communities corresponding to those dense agglomerated clusters, which are relevant in the neurobiological context and are potentially consistent with coherent neural activities.

\begin{itemize}

\item \textbf{Criterion 1:} The community structure of the graph was identified. The average degree of the graph was calculated. Were considered as communities highly integrated, those communities whose sub-graphs respective to the vertexes of the community presented a value superior to or equal to 70\% of the mean connectivity of the complete graph.

 \item \textbf{Criterion 2:} The community structure of the graph was identified. We considered as highly connected the communities that presented eight or more vertexes. It was empirically observed from the analysis of the community structure of several graphs estimated that highly connected communities presented a larger number of vertexes.
 
\end{itemize}

The two criteria revealed to be equivalent\footnote{It was verified empirically from the observations, the validity, and equivalence of both criteria. Those criteria may not be valid for graphs of other nature, size, or community structures detected by different algorithms.}, presenting similar results. All the figures and the results of this study involved the use of the \textbf{criterion 2}, which was chosen because it was simpler.


\section{Results}

\textbf{Notes}: 

\textit{ 
\textbf{1)} The community structure presented in this section corresponds to communities found by the algorithm Walktrap Community \citep{pons2006computing} that satisfied the \textbf{criterion 2}, described in methods \textbf{Subsection V}. Vertexes belonging to communities that did not fulfill this criterion were represented in white in the pictures.}\\

\textit{\textbf{2)} The communities presented in the figures correspond to regions highly integrated functionally, with the vertexes being densely connected to themselves and less densely connected to other vertexes of the network. They indicate the presence of highly integrated and segregated neural processes and not simply ``active'' or ``inactive'' areas.}

\subsection{Community Structure}

In the experiment, it was possible to observe the community structure of functional brain networks. The results have provided experimental evidence demonstrating that functional brain activities are modularly structured. This organization character was observed on the five physiological frequency bands analyzed in both conditions, awake and anesthetized.

The community structure of functional brain networks was not shown to be static; it was revealed to be dynamic, constantly presenting changes over time. The communities varied in number, size, and anatomical areas involved. It was possible to observe that vertexes belonging to the same module tended to be physically close. The functional clusters, in general, presented continuous shapes, and their spatial location seemed to have a consistent relation with anatomical areas and divisions.

Despite the dynamic nature, it was observed that some patterns appeared with a higher likelihood. Each physiological frequency band and state, awake or anesthetized, presented a singular and characteristic community structure\footnote{Except in Delta frequency band, where no visible alterations occurred before and after anesthesia.}. Those results revealed that the modular organization character of neural activities was not degraded to the point of extinguishing during general anesthesia; on the contrary, ``richness'' was observed in the complexity and dynamics of the community structure during general anesthesia.

\subsection{Community Structure in Wake and Anesthetized States}

Except at lower frequencies (Delta band \mbox{0-4Hz}), it was possible to verify the existence of considerable changes regarding the awake and anesthetized states. The most prominent alterations in the community structure concerned the areas in which the clusters were located.
The awake state was characterized by communities involving large frontal lobe areas and a high tendency to occupy most of the parietal regions. Large communities involving concomitantly frontal and parietal areas were also frequent in awake conditions. Another remarkable characteristic that contrasted the structure of the anesthetized state was the absence of communities in the occipital lobe and posterior temporal areas\footnote{The absence of communities was observed most of the time when the monkey was blindfolded.}.\\

Predominantly, the state of anesthesia was characterized by a reduction in the presence of communities involving frontal and parietal areas, being evident an apparent reduction in the occupancy of communities on the frontal lobe, in opposition to what was observed during the awake state. Another remarkable characteristic observed during anesthesia was the massive appearance of large communities at the occipital lobe, which frequently extended to temporal regions and also to the motor and somatosensory cortices, involving the central sulcus and nearby areas.

Comparing the awake and anesthetized states were observed a reduction in the presence and coverage of communities in the secondary associative cortex and an increase in the occupancy at the primary motor and sensory cortices.

\subsubsection{Delta (0-4HZ)}
The community structure in the Delta band (\hyperlink{FIGURE2}{$Figure \cdot 2$}) was characterized by the presence of clusters on all or almost all the cortical surfaces on which ECoG electrodes were positioned. No particular region characterized by the absence of communities was found\footnote{There were moments in which some regions did not present communities, however, the location of those regions varied, not characterizing a recurrent pattern.}. In general, three to five communities were observed at each time. In addition, several modules embraced quite large cortical areas, and some anatomical divisions seemed to constrain the boundaries of the clusters.

\end{multicols}

\begin{figure*}[ht]
  \includegraphics[width=\textwidth,height=8.75cm]{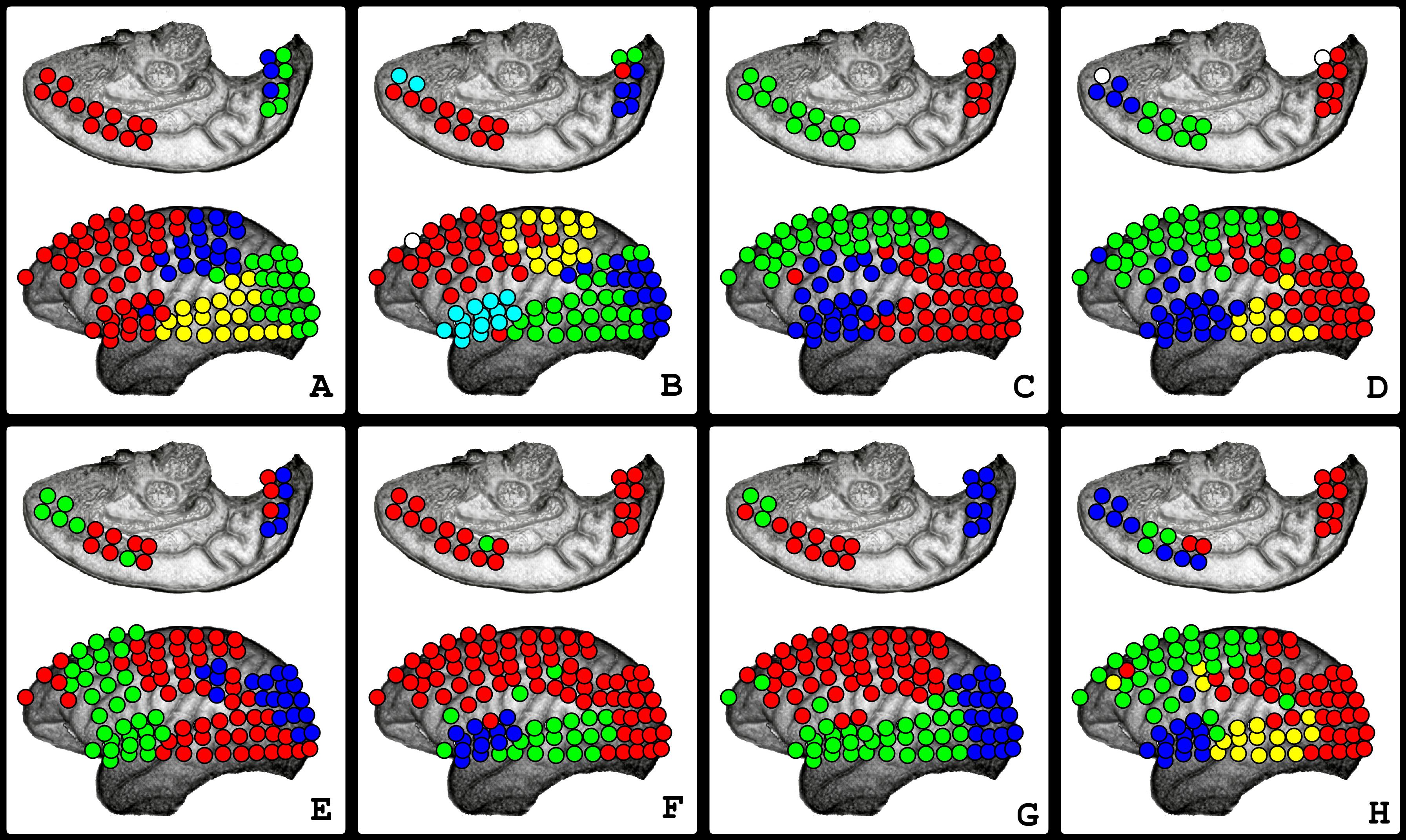}
  \caption{
\textbf{Community structure of the Delta frequency band (0-4Hz)}, sub-figures A, B, C, and D correspond to the most representative community structures found during awake conditions; sub-figures E, F, G, and H correspond to the most representative community structures found after the administration of the anesthetics. The colors discretize different communities, each represented by a different color. The white color indicates the absence of communities. Distinct colors do not possess specific meaning besides indicating the presence and coverage of functional clusters.
}
\hypertarget{FIGURE2}{}
\end{figure*}

\begin{multicols}{2}

The superior temporal sulcus typically delimited the extension of some communities (see \hyperlink{FIGURE2}{$Figure \cdot 2$, \mbox{\textit{Sub-Figures A, B, C, D, E, F, and H}}}); some communities seemed to respect brain lobes anatomy divisions (see \hyperlink{FIGURE2}{$Figure \cdot 2$} A).  

No specific patterns distinguished the awake state (\hyperlink{FIGURE2}{$Figure \cdot 2$, \mbox{\textit{Sub-Figures A - D}}}) from the state of anesthesia (\hyperlink{FIGURE2}{$Figure \cdot 2$, \mbox{\textit{Sub-Figures E - K}}}). No remarkable change was observed in the dynamics of the community structure during the anesthetic induction process.

\enlargethispage{1.5\baselineskip}

\subsubsection{Theta (4-8Hz)}

One of the characteristics of the Theta band was the absence of functional clusters in some regions, in both awake and anesthetized states. It was possible to verify apparent alterations in the community structure before and after administering the anesthetics.

In awake conditions (\hyperlink{FIGURE3}{$Figure \cdot 3$, \mbox{\textit{Sub-Figures A - D}}}), it was possible to observe a tendency in the presence of clusters to involve the entire frontal and parietal lobes most of the time, communities involving both lobes were also frequent. There was a tendency of an absence of communities in quite large areas at the occipital and temporal lobes\footnote{While the macaque was blindfolded, large functional clusters on those regions were not frequent.} (see \hyperlink{FIGURE3}{$Figure \cdot 3$, \mbox{\textit{Sub-Figures A, B, and D}}}).

The community structure found during general anesthesia (\hyperlink{FIGURE3}{$Figure \cdot 3$, {\textit{Sub-\mbox{Figures E - H}}}}) was characterized by the absence of clusters in considerable areas of the frontal lobe. Clusters that spread to the inferior and medial regions of the frontal lobe seemed to be the most affected. Communities still involved superior areas of the frontal and parietal lobes. Another characteristic that contrasts with what was observed in the awake state was the presence of communities involving all of the occipital lobe and large areas of the temporal lobe (see \hyperlink{FIGURE3}{$Figure \cdot 3$, \mbox{\textit{Sub-Figures E - H}}}).

\end{multicols}

\begin{figure*}[!ht]
  \includegraphics[width=\textwidth,height=8.75cm]{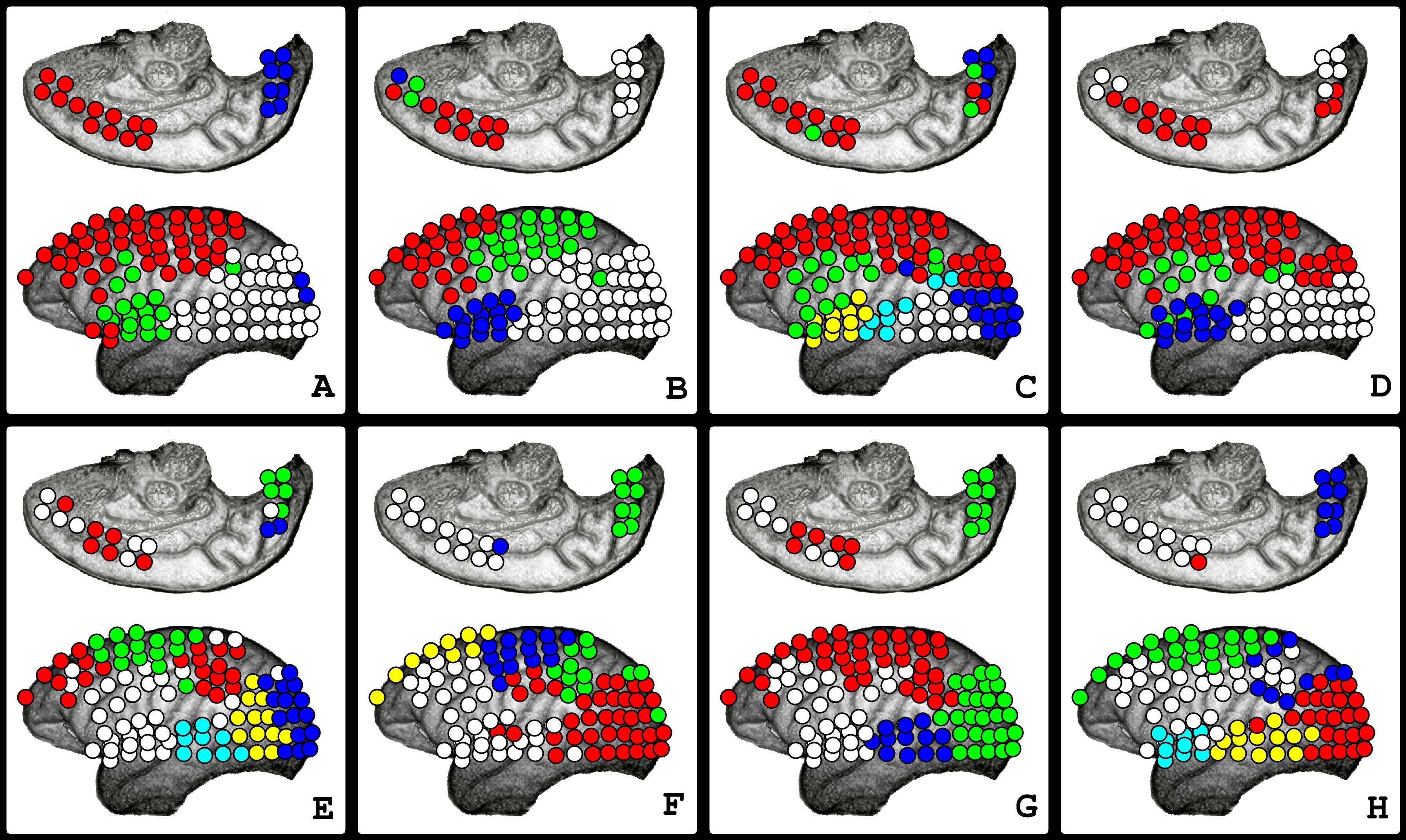}
  \caption{\textbf{Community structure of the Theta frequency band (4-8Hz)}, sub-figures A, B, C, and D correspond to the most representative community structures found during awake conditions; sub-figures E, F, G, and H correspond to the most representative community structures found after the administration of the anesthetics. The colors discretize different communities, each represented by a different color. The white color indicates the absence of communities. Distinct colors do not possess specific meaning besides indicating the presence and coverage of functional clusters.
}
\hypertarget{FIGURE3}{}
\end{figure*}

\begin{multicols}{2}

\subsubsection{Alpha (8-12Hz)}

On the alpha frequency band, not all cortical regions analyzed contained clusters in both awake and anesthesia conditions. 

An evident distinction in the community structure was verified between awake and general anesthesia conditions.

The awake state (\hyperlink{FIGURE4}{$Figure \cdot 4$, \mbox{\textit{Sub-Figures A - D}}}) was characterized by communities involving entirely frontal and parietal lobes most of the time. Clusters were also frequently found in the anterior and medial parts of the temporal lobe. Communities were not frequently observed in the occipital lobe and posterior temporal areas (see \hyperlink{FIGURE4}{$Figure \cdot 4$, \mbox{\textit{Sub-Figures A, B, and D}}}).

The community structure of the anesthetized state (\hyperlink{FIGURE4}{$Figure \cdot 4$, \mbox{\textit{Sub-Figures E - H}}}) was characterized by a considerable reduction in the occupancy of communities at the frontoparietal secondary associative cortex. Another characteristic of that state was the massive presence of clusters at the occipital lobe and on the primary motor and somatosensory cortex (central sulcus and areas nearby) (see \hyperlink{FIGURE4}{$Figure \cdot 4$, \mbox{\textit{Sub-Figures E - H}}}); the location and shape of some clusters captured almost perfectly the anatomy of the central sulcus.

\end{multicols}

\begin{figure*}[!ht]
  \includegraphics[width=\textwidth,height=8.75cm]{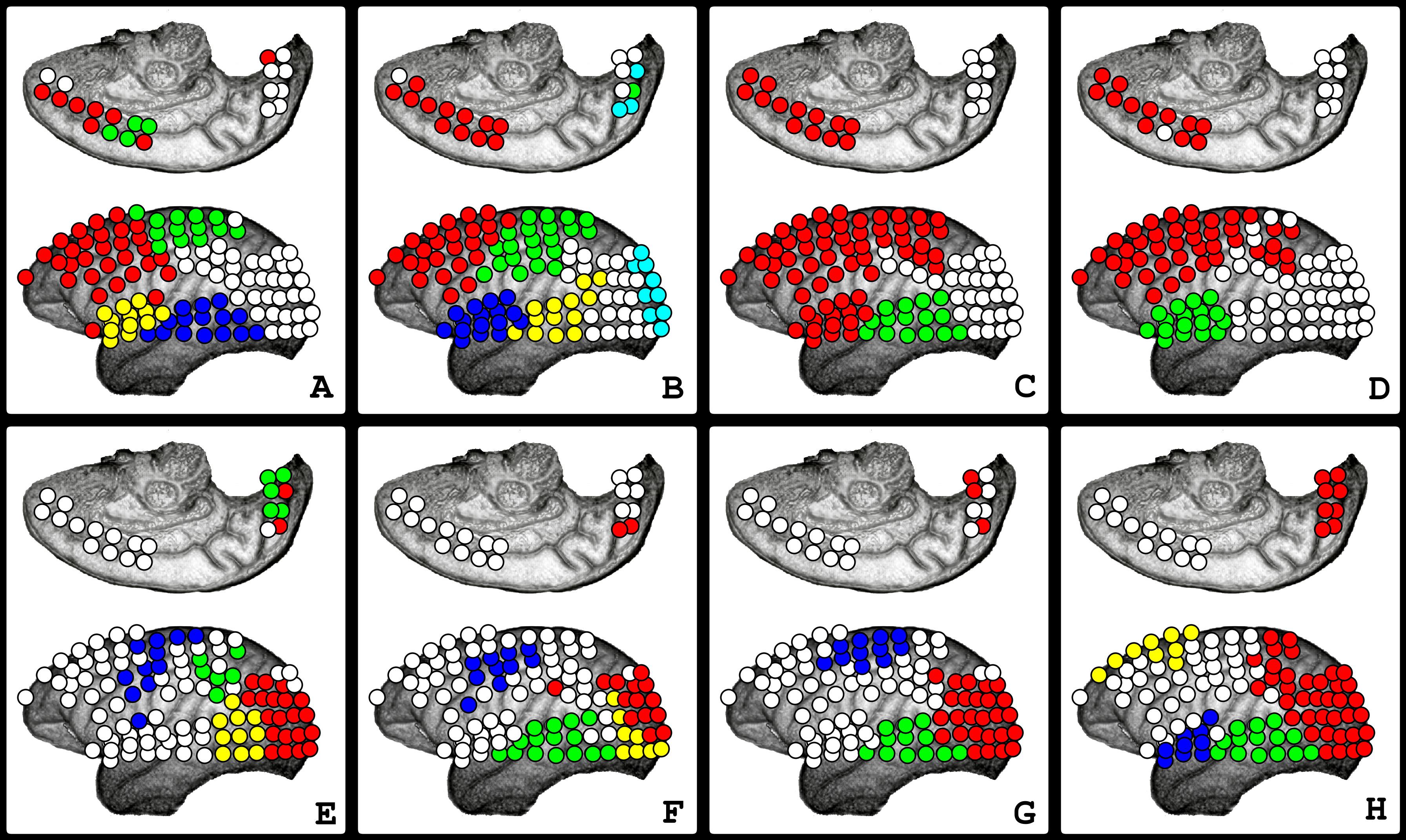}
  \caption{\textbf{Community structure of the Alpha frequency band (8-12Hz)}, sub-figures A, B, C, and D correspond to the most representative community structures found during awake conditions; sub-figures E, F, G, and H correspond to the most representative community structures found after administering the anesthetics. The colors discretize different communities, each represented by a different color. The white color indicates the absence of communities. Distinct colors do not possess specific meaning besides indicating the presence and coverage of functional clusters.
}
\hypertarget{FIGURE4}{}
\end{figure*}

\begin{multicols}{2}

\subsubsection{Beta (13-30Hz)}

On the Beta band, not all cortical regions in which the electrodes were positioned presented communities. After administering the anesthetics, considerable alterations were observed (see \hyperlink{FIGURE5}{$Figure \cdot 5$}).

The awake state (\hyperlink{FIGURE5}{$Figure \cdot 5$, \mbox{\textit{Sub-Figures A - D}}}) presented, on average, three to four communities. Clusters that involved large areas of the frontal and parietal lobes were frequently observed. It was also verified that regions of the occipital and temporal lobes (mainly posterior areas), most of the time, did not present communities (see \hyperlink{FIGURE5}{$Figure \cdot 5$, \mbox{\textit{Sub-Figures A - D}}}).

During anesthesia (\hyperlink{FIGURE5}{$Figure \cdot 5$,  \textit{Sub-Figures\mbox{ E - H}}}), a considerable reduction in the occupancy of clusters was observed mainly in regions of the frontal and parietal lobes, except in the central sulcus and areas nearby, in which communities were frequently observed. In opposition to what was observed during the awake state, a massive presence of communities in the occipital lobe and posterior parts of the temporal lobe was verified (see \hyperlink{FIGURE5}{$Figure \cdot 5$, \mbox{\textit{Sub-Figures E - H}}}).

\end{multicols}

\begin{figure*}[!ht]
  \includegraphics[width=\textwidth,height=8.75cm]{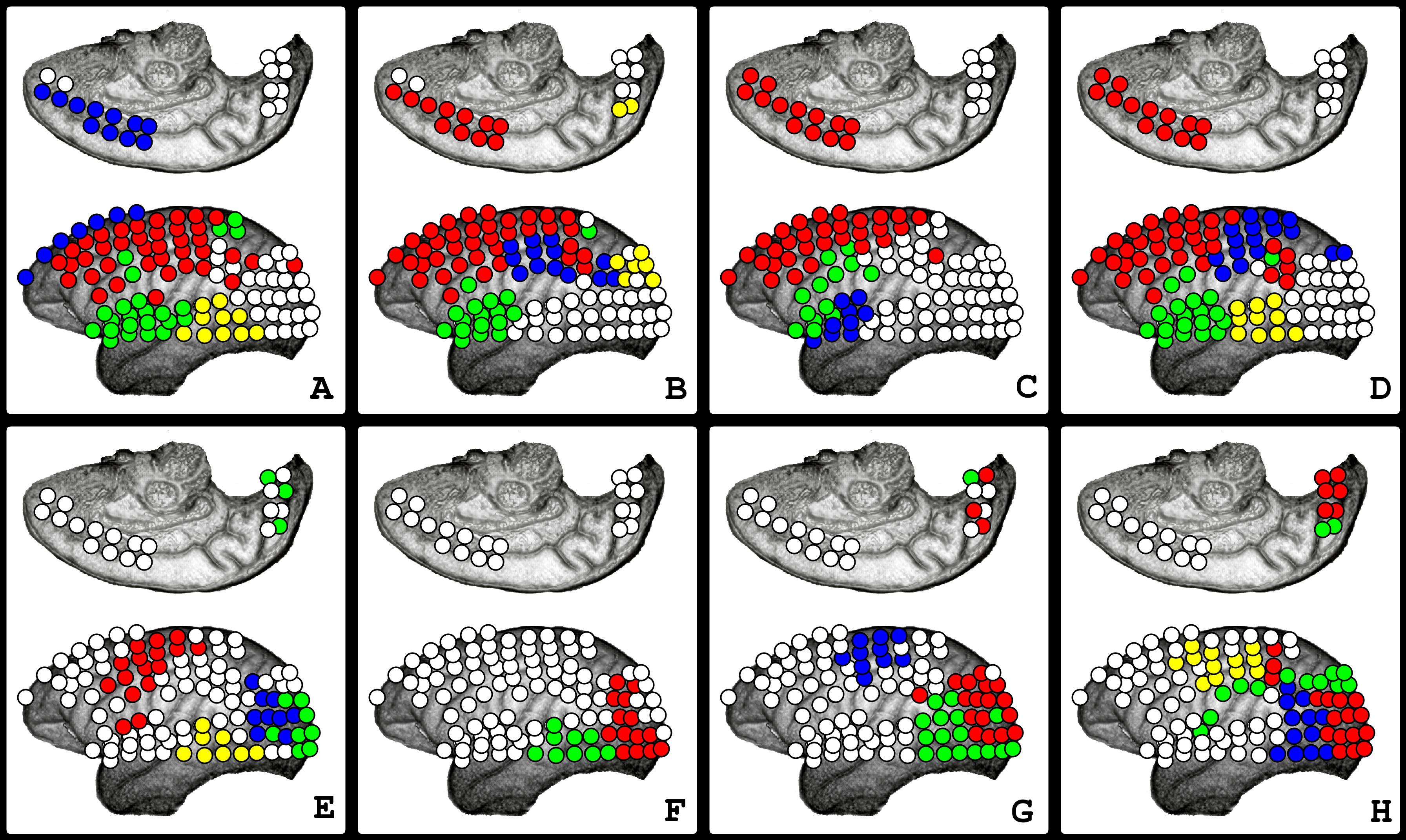}
  \caption{\textbf{Community structure of the Beta frequency band (13-30Hz)}, sub-figures A, B, C, and D correspond to the most representative community structures found during awake conditions; sub-figures E, F, G, and H correspond to the most representative community structures found after the administration of the anesthetics. The colors discretize different communities, each represented by a different color. The white color indicates the absence of communities. Distinct colors do not possess specific meaning besides indicating the presence and coverage of functional clusters.
}
\hypertarget{FIGURE5}{}
\end{figure*}

\begin{multicols}{2}

\subsubsection{Gamma (25-100Hz)}

On the Gamma frequency band, both states had regions that did not present communities. Visible changes were verified after administering the anesthetics (see \hyperlink{FIGURE6}{$Figure \cdot 6$}).

The awake state (\hyperlink{FIGURE6}{$Figure \cdot 6$, \mbox{\textit{Sub-Figures A - D}}}) was characterized by communities involving large portions of the frontal and parietal lobes, except in some instants when the spread of the functional clusters was naturally reduced. The absence of communities in large areas of the occipital and temporal lobes was observed (see \hyperlink{FIGURE6}{$Figure \cdot 6$, \mbox{\textit{Sub-Figures A - E}}}).

The induced state of anesthesia (\hyperlink{FIGURE6}{$Figure \cdot 6$, \mbox{\textit{Sub-Figures E - H}}}) was characterized by a reduction in the occupancy of communities on the frontal and parietal lobes, except for the presence of communities on the superior regions of the frontal lobe and medial frontal walls. In addition, a considerable presence of communities at the occipital and temporal lobes was observed (see \hyperlink{FIGURE6}{$Figure \cdot 6$, \mbox{\textit{Sub-Figures E - H}}}).

\end{multicols}

\begin{figure*}[!ht]
  \includegraphics[width=\textwidth,height=8.75cm]{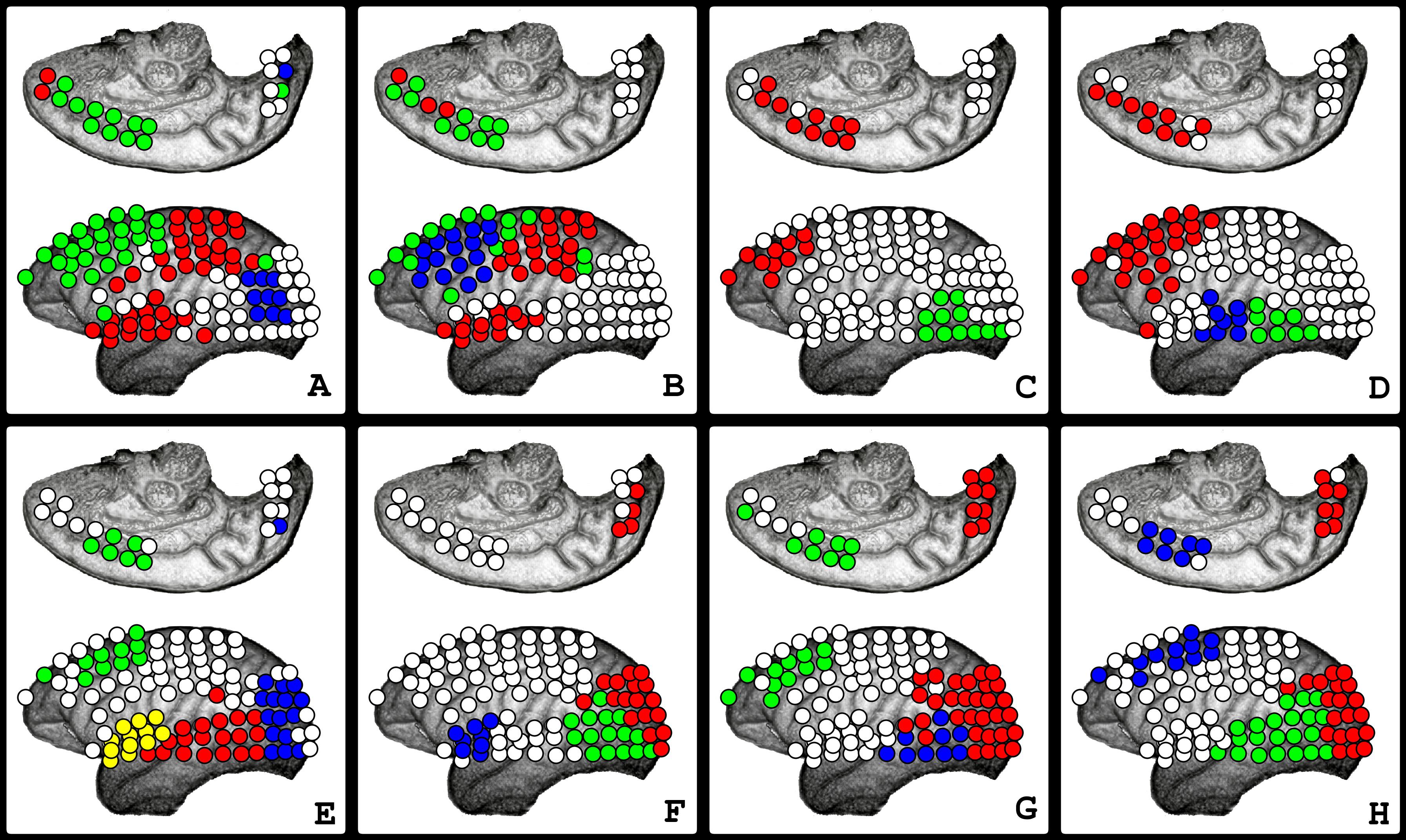}
  \caption{\textbf{Community structure of the Gamma frequency band (25-100Hz)}, sub-figures A, B, C, and D correspond to the most representative community structures found during awake conditions; sub-figures E, F, G, and H correspond to the most representative community structures found after the administration of the anesthetics. The colors discretize different communities, each represented by a different color. The white color indicates the absence of communities. Distinct colors do not possess specific meaning besides indicating the presence and coverage of functional clusters.
}
\hypertarget{FIGURE6}{}
\end{figure*}

\begin{multicols}{2}

\subsection{Transitions}

It was possible to observe that the macaque large-scale functional brain network community structure does present changes over time. Besides being dynamic, on each frequency band, two main patterns were observed along the experiment\footnote{Except on the Delta frequency band, which did not present visible alterations after the administration of the anesthetics.}, the first respective to the awake state, and the second to the induced-state of anesthesia.
From the observation of the community structure of the networks estimated sequentially over time (at every five seconds), it was possible to verify transitions between the two states. The first was respective to the anesthetic induction (see \hyperlink{FIGURE7}{$Figure \cdot 7$}), and the second to the recovering process (see \hyperlink{FIGURE8}{$Figure \cdot 8$}). It was possible to observe how the transitions happened, how long they lasted, and how long after the administration of the drugs, changes started to occur. 

\newpage

\textbf{Note:}\textit{ The transitions during anesthetic induction and recovery were clearly noticeable on the Alpha, Theta, Beta, and Gamma frequency bands. The present manuscript presents only the transitions that occurred on the Beta band. }

\subsubsection*{Anesthetic Induction}

\begin{figure*}[!ht]
\includegraphics[width=\textwidth]{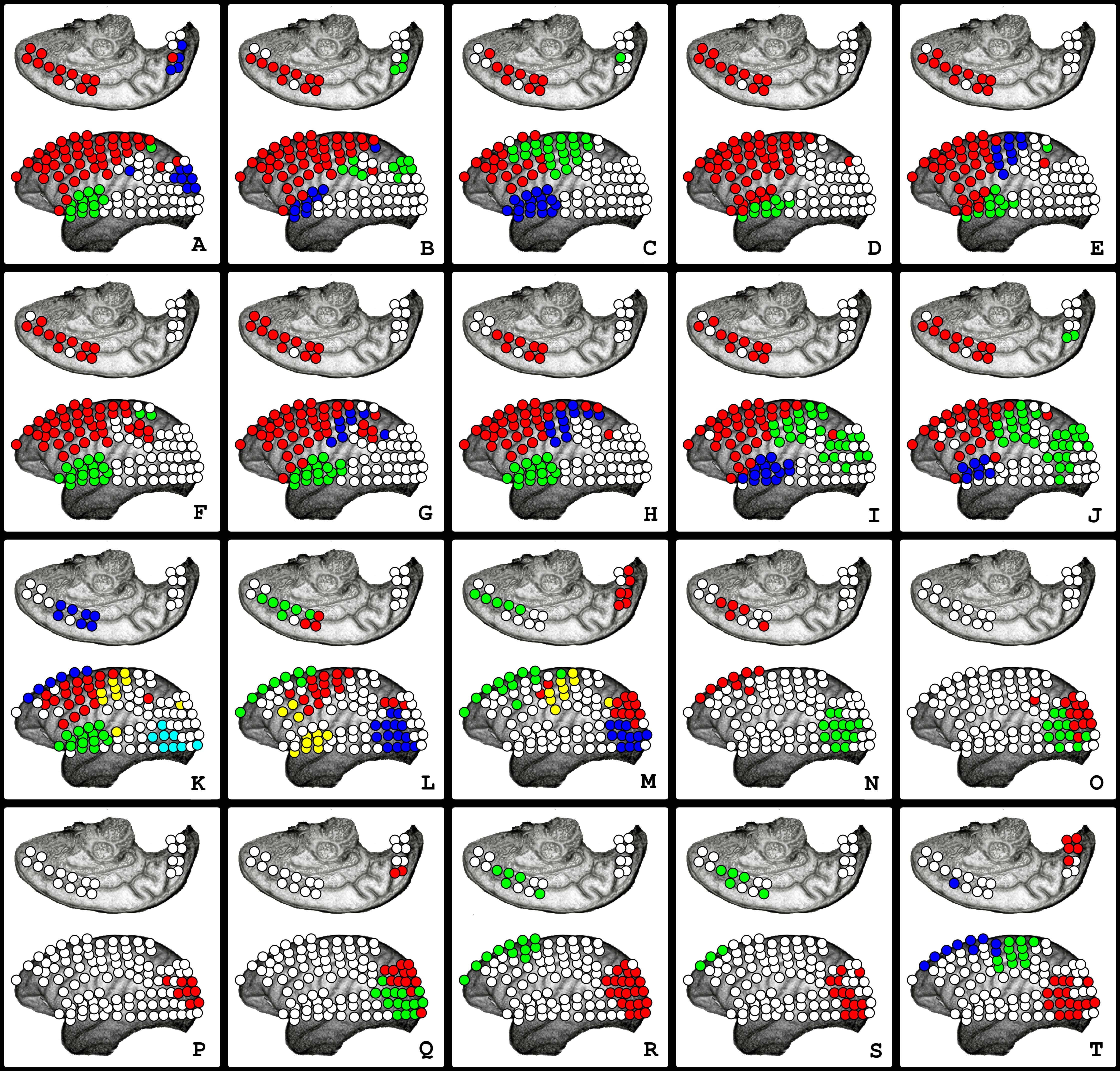}
\caption{\textbf{Community structure}. The transition between awake and anesthetized state during the anesthetic induction. Frequency band Beta (13-30Hz). The sub-figures represent the community structure sequentially inferred along time, being 5 seconds the time interval between each frame and its subsequent. The sub-figure A starts approximately one minute after the administration of the anesthetics. The colors discretize different communities, each represented by a different color. The white color indicates the absence of communities. Distinct colors do not possess specific meaning besides indicating the presence and coverage of the functional clusters.
}
\hypertarget{FIGURE7}{}
\end{figure*}

In the Beta band, the beginning of the transition started about one and a half minutes after administering the anesthetics (see the beginning of the transition in \hyperlink{FIGURE7}{$Figure \cdot 7$, \mbox{\textit{Sub-Figure I}}}). The transition was quite fast, taking about 30 to 35 seconds (see \hyperlink{FIGURE7}{$Figure \cdot 7$, \mbox{\textit{Sub-Figures I - N}}}). The appearance of communities in the occipital region, along with the reduction of the spread of communities in frontal and parietal areas, was verified (observe \hyperlink{FIGURE7}{$Figure \cdot 7$, \mbox{\textit{Sub-Figures I - N}}}), and it was observed that the transition between the two states happened through intermediary\footnote{Community structures possessing characteristics of both states.} community structures, shifting from one characteristic community structure pattern to the other.

It was possible to verify the transition during the recovery process (see \hyperlink{FIGURE8}{$Figure \cdot 8$}). After administering the antagonist of Medetomidine, Atipamezole, noticeable alterations started within about two minutes.

Different from the anesthetic induction, in which as soon as the first alterations occurred, the transition happened rapidly and at once. During the recovery process, the transition was slower and more complex. The transition was not given at once; during the recovery, there were moments in which functional clusters occupied large cortical areas (see \hyperlink{FIGURE8}{$Figure \cdot 8$, \mbox{\textit{Sub-Figures A, C, L, and R}}}). Periods in which the presence of communities was reduced substantially (see \hyperlink{FIGURE8}{$Figure \cdot 8$, \mbox{\textit{Sub-Figures M, P, and Q}}}) and times when patterns resembling the state of anesthesia were observed (see \hyperlink{FIGURE8}{$Figure \cdot 8$, \mbox{\textit{Sub-Figures D - J}}}). Right after the occurrence of the first alterations, community structures characteristic of the awake state were observed (see \hyperlink{FIGURE8}{$Figure \cdot 8$, \mbox{\textit{Sub-Figure B}}}) without a gradual transition. Some minutes after the structure of the awake state seemed established, it was possible to observe some moments in which patterns of the anesthetized state still appeared again, revealing the existence of an oscillatory shift between the two states.

\end{multicols}

\begin{figure*}[p]
\includegraphics[width=1\textwidth]{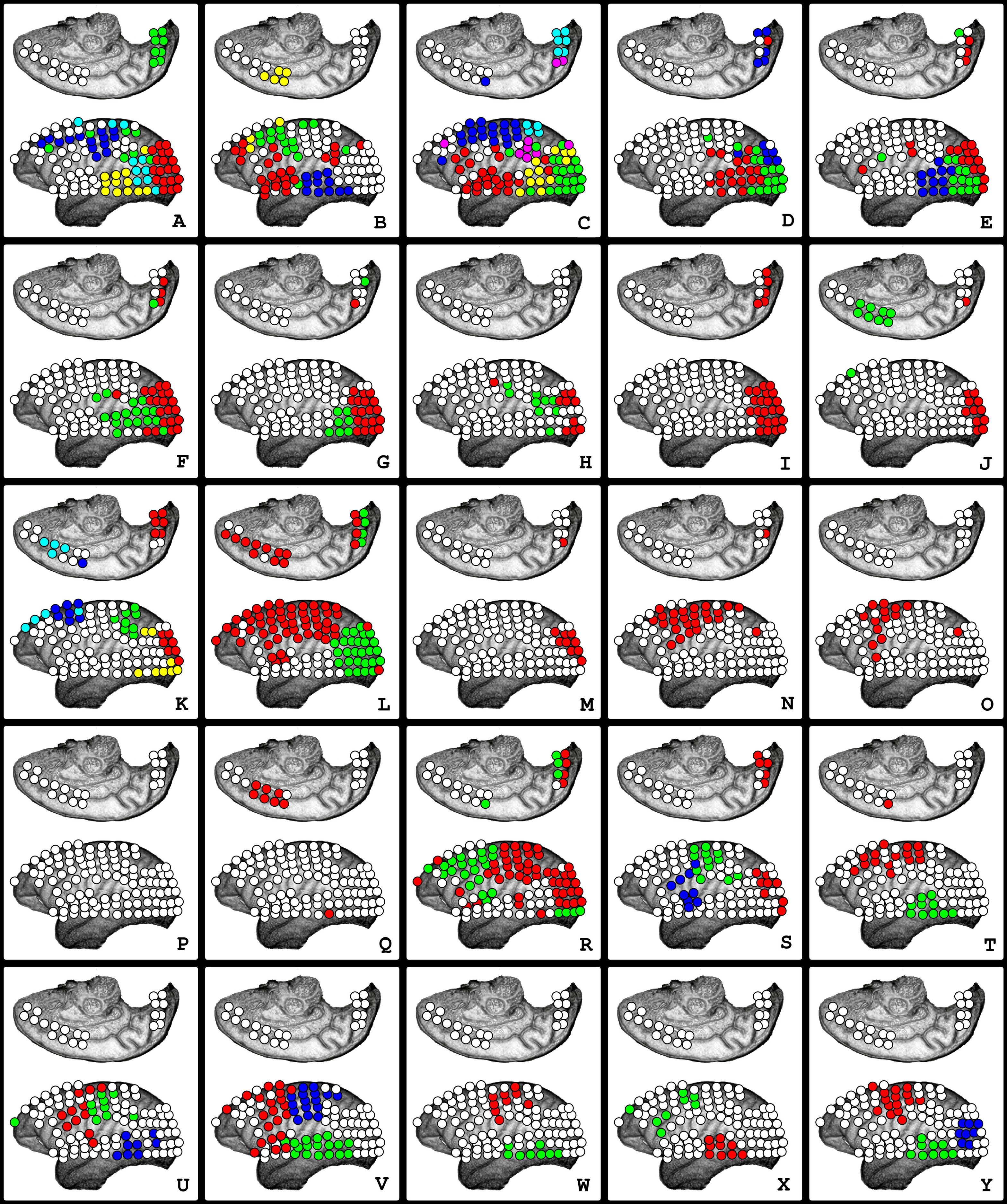}
\caption{
\textbf{Community structure}. The transition between the anesthetized and awake states during the recovery process. Frequency band Beta (13-30Hz). The sub-figures represent the community structure sequentially inferred along time, being 5 seconds the time interval between each frame and its subsequent. The sub-figure A starts approximately two minutes after the administration of the antagonist Atipamezole. The colors discretize different communities, each represented by a different color. The white color indicates the absence of communities. Distinct colors do not possess specific meaning besides indicating the presence and coverage of the functional clusters. }
\hypertarget{FIGURE8}{}
\end{figure*}

\begin{multicols}{2}

 \vspace{-1.5\baselineskip}
\section{Discussion}

From the experimental setup and data analysis methodology, it was possible to contemplate the structure and dynamics of the segregated and integrated neural processes, providing an invaluable opportunity to observe the working brain.
The notable presence of communities on all frequency bands analyzed constitutes experimental evidence that large-scale functional neural networks are modularly structured, revealing that over time certain cortical regions are highly integrated, functionally interacting much more with themselves than with other areas of the cortex. Those results are in accordance with predictions made by several neuroscientists stating that brain activities should be structured this way. Once this architecture promotes balance between specialized areas, minimizes wiring and energy costs \citep{bullmore2012economy} and supports both segregated and integrated information processing \citep{sporns2000theoretical,tononi1994measure,
meunier2010modular,gallos2012small}, which are thought to be two major organizational principles of the cerebral cortex \citep{zeki1978functional,zeki1988functional,
tononi1994measure,tononi1998consciousness,
friston2002beyond,friston2005models,
friston2009modalities}, suggests that this form of organization of functional brain activities should be expected, agreeing with the experimental results found in the present research.

\enlargethispage{1.0\baselineskip}

\subsubsection*{Structure and Patterns}

The functional clusters observed in this experiment presented well-defined patterns, frequently involving large areas of the cortex, and contained vertexes (anatomical areas) located geographically closer to each other, typically presenting continuous shapes. The fact that well-defined patterns respecting anatomical boundaries were frequently observed suggests a strong relationship regarding the location and extension of functional clusters over specific cortical anatomical areas. These results are in accordance with experimental evidence based on the community structure of connection datasets of cortical thickness correlation \citep{chen2008revealing} and cortical connectivity of macaque and cat \citep{hilgetag2000anatomical}, that demonstrate the existence of a significant overlap between anatomical networks modules and functional systems in the cortex.

The observation of functional clusters involving large cortical areas brings evidence that neural processes and activities may involve the joint participation of several specific cortical regions, most probably each performing a distinct ``sub-function'' in this integrated process. The continuous extension and the geographical proximity between the vertexes belonging to the same clusters also indicate that the structure of functional neural activities may follow some optimization-economy principles. There are also theoretical predictions that functional clusters should be structured this way. Anatomically speaking, spatially close brain regions are more likely to be connected once interactions between physically distant brain regions supposedly involve longer axonal projections and high energy costs. This way, functional clusters should involve physically close anatomical areas \citep{bullmore2012economy}, being the experimental observations in agreement with the theoretical predictions.

\subsubsection*{Dynamics}

The dynamics of those activities and neural processes were verified through the analysis of the community structure inferred sequentially over short time intervals. It was possible to contemplate a wide variety of patterns, numbers, and coverage of the functional clusters. Being revealed that the working brain is highly dynamic and active. Those experimental observations are per statements by McIntosh that brain processes may be accomplished by network interactions that rapidly reconfigure involving several brain regions, leading to dynamic changes in neural context \citep{mcintosh1999mapping,mcintosh2000towards,
mcintosh2008large}. 

 The experimental results strongly indicate that during the resting state\footnote{The resting state is a state in which the subject is not performing an explicit task that possibly involves cognitive demands.}, and even during general anesthesia, there might be a set of activities happening in the brain that seemed to be performing much more than homeostasis. Furthermore, despite the brain having demonstrated as highly dynamic and active, in both states analyzed, the existence and recurrence of some similar patterns were observed, and some groups of anatomical areas were most of the time included. Those results bring evidence that possibly the same set of neural processes and activities may be constantly repeating in different instants of time.

\subsubsection*{Transitions}

The analysis of the community structure over short intervals of time revealed the existence of transitions in both the anesthetic induction and recovery processes. Those results are very interesting according to the perspective of the \textit{Modern Network Science} \citep{strogatz2001exploring,newman2003structure} and \textit{Systems Neuroscience} \citep{van200523} which once highlighted a phase transition in the structural properties of large-scale functional brain networks and a change in the physiological state of the organism. Furthermore, as the functional clusters indicate the presence of highly integrated and segregated neural processes and alterations in the areas involved, shapes, extensions, and dynamics of those processes at the onset of the anesthetic induction, potentiality relates some features of the functional clusters with neural correlates of consciousness.

From the analysis of (\hyperlink{FIGURE7}{$Figure \cdot 7$, - \textit{Anesthetic Induction}}), it is possible to observe that the transition in the community structure happened within about 30 to 35 seconds, which is in agreement with reports by Emery Brown, that under the anesthetic induction, the transition to the state of unconsciousness occurs in a fast way, requiring about only a few tenths of seconds \citep{brown2011general}. However, the results of this study also indicate that the recovery process is more complex and slower, taking some minutes for the patterns of the awake state to re-establish.

No straightforward alterations were observed at the loss of consciousness point (LOC), which was registered about ten minutes after administering the anesthetics when the animal no longer reacted to external stimuli such as touching the nostrils or opening the hands. The awake state probably ended when alterations in the community structure of the networks occurred, which happened less than two minutes after the anesthetic injection. However, after that time, the monkey was still able to present involuntary reactions to stimuli, being in an intermediary state between deep sedation and general anesthesia, probably in a similar condition as the pharmacological effects of the administration of Ketamine in humans \citep{bergman1999ketamine}.

\subsubsection*{Structures of the Awake and the Anesthetized States}

During awake conditions, the most remarkable feature of the community structure of functional brain networks was the presence of communities involving large areas of the frontal and parietal lobes. Functional clusters involving both regions were also frequent in that state. These observations suggest that the awake condition may be dependent on highly integrated neural processes involving these same areas. These suppositions are in accordance with the predictions of Tononi and Edelman that conscious experiences would typically be related to highly integrated and differentiated neural processes involving posterior corticothalamic regions involved in perceptual categorization interacting functionally with anterior regions involved in concept formation, value-related memory, and planning \citep{tononi1998consciousness,
edelman2013consciousness}. \\

The main areas affected by the administration of the Ketamine-Medetomidine cocktail were the secondary associative cortex of the frontal and parietal lobes. This experimental evidence is consistent with several scientific studies and reports relating the same regions to neural correlates of consciousness. There is experimental evidence that the administration of the general anesthetic Propofol induces a significant decrease in the blood flow in areas of the pre-cuneus, cuneus, and posterior cingulate cortex \citep{fiset1999brain}, with the inactivation of the posterior medial part being associated with the loss of consciousness \citep{kaisti2002effects}. Damasio and Dolan reported that injuries affecting the same cortical areas are related to severe impairments in cognition and consciousness \citep{damasio1999feeling}. Laureys highlights that frontoparietal regions are preferentially deactivated in human patients in the vegetative state \citep{laureys2004brain}, and the loss of consciousness in those patients is associated with the functional disconnectivity between frontal and parietal regions \citep{laureys1999impaired}.


\subsubsection*{Community Structure and Vertexes Degree}

To verify if there was a direct relation between the number of connections of the networks vertexes and the presence of clusters in different cerebral regions, the community structure and the degree of the vertexes of the networks \citep{padovani2016characterization} in different cerebral regions were jointly analyzed.
It was observed that in frontal, parietal, and temporal regions, a significant decrease in the presence and occupancy of functional clusters happened; meanwhile, a substantial reduction in functional connectivity of the vertexes occurred \citep{padovani2016characterization}. This observation indicates that the anesthetic agents affected the high functional integration observed in those areas during the awake state and consequently impacted the occurrence of neural processes highly integrated with characteristics of functional clusters present in those regions while the monkey was awake.

However, it was possible to verify that the presence or absence of communities was not associated only with the increase or decrease of the degree of the vertexes in the corresponding regions. Instead, structural rearrangements in how interactions between the areas were established (structural aspects of the graph) were also involved in the presence or absence of communities.
Analyzing the figure ({$Figure \cdot 3$}, \textit{Sub-Figure D, pg.10}) of \citep{padovani2016characterization}\footnote{Study involving the same database.}, it is possible to observe that in the first five minutes after the first effects of the anesthetics were observed, the average degree respective to the sub-graphs of the occipital lobe diminished, reaching smaller values than those observed during awake conditions (eyes closed). However, during that time, even with a decline in the degree of the vertexes,  communities dramatically appeared in that region, indicating that structural factors were also involved in the establishment of communities.

After blindfolding the macaque, it was verified that the functional networks had changed considerably \citep{padovani2016characterization}; however, no expressive changes in the community structure were observed, except for a decrease in the presence of clusters in occipital areas, which happened even though the average connectivity of the sub-graphs respective to that region had almost doubled in mean values\footnote{ Compared to the time in which the monkey was with eyes opened.} \citep{padovani2016characterization}. Except for those alterations observed at the occipital lobe, the community structure in the same physiological state proved to be quite robust against changes in the several properties of the networks that happened after the placement of the blindfold \citep{padovani2016characterization}.

\subsection{Assumptions and Hypotheses}

This section presents some hypotheses and assumptions with the objective of explaining the results observed experimentally. 

\subsubsection*{Main Changes Observed}

The most noticeable changes and phenomena observed in the neural activities of the macaque during the Ketamine-Medetomidine-induced state of anesthesia were:

\begin{enumerate}

\item Transition to a highly specific and complex state, distinct from the state that existed when the animal was awake.

\item Expressive reduction of the functional connectivity \citep{padovani2016characterization}, decrease in coverage and number of functional clusters mainly in regions correspondent to the secondary associative cortex.

\item The substantial appearance of functional clusters mainly in areas of the primary sensory and motor cortices, also extending to nearby regions.

\end{enumerate}

Given the experimental observations, some hypotheses were formulated to explain, at least in part, the phenomena and changes observed.\\

During the Ketamine-Medetomidine-induced state of anesthesia, it was observed that regions of the secondary associative cortex were affected. In contrast, regions relative to the primary sensory and motor cortices were favored. Those results suggest that the differences between the primary and secondary associative cortices were related to the observed phenomena. It is necessary to comprehend some principles of the cerebral cortex organization to understand how these differences might be involved with the results.
The cortex comprehends the most external layer of the brain, possessing only a few millimeters of thickness. Along the entire cerebral mantle, it assumes a laminar organization structure, being possible to be discretized into subdivisions composed of distinct populations of cells of different densities, sizes, and shapes \citep{purves2012neuroscience}. Schematically, the microcircuits found on this laminar structure perform three main functions: the amplification of inputs and signals received, the computation performed by operations realized mainly on the superficial layers, and the communication that involves the transmission of information from one layer to another, and also from other regions of the cortex. Besides a global similarity, different specialized cortical areas possess distinctions in this laminar organization \citep{brodmann1908beitrage}, presenting cells with distinct morphology, cellular connectivity (distinct neuronal circuits), and different sources of inputs and destinations of outputs. One of the main differences between the primary sensory and motor cortex (regions predominantly activated) and the secondary associative cortex (regions most affected) resides in the source of inputs and destination of outputs of these regions. The primary sensory and motor cortices are related to the primary processing of sensory and motor information, with the thalamus being its main source of inputs\footnote{The primary sensory cortex receives thalamic information related directly to the peripheral sensory organs. The motor cortex receives inputs from the thalamic nuclei related to basal ganglia and cerebellum.}. The secondary associative cortex performs a function of association, involved with the integration of the information that has been processed sequentially in primary and higher-order divisions of the sensory and motor cortex, being its primary source of input to other regions of the cortex and not the thalamus\footnote{The secondary associative cortex also receives inputs from the thalamus, but its\textbf{ \underline{primary}} source of inputs comes from other regions of the cortex.}.

\subsubsection*{Assumptions: Affected Areas}

An expressive reduction in functional connectivity \citep{padovani2016characterization} and the presence of functional clusters in areas respective to the secondary associative cortex were observed. It is not easy to propose a mechanism or a hypothesis explaining those experimental findings. However, it is known that Ketamine is a drug able to modulate the activity of a specific ion channel protein. Consequently, it may affect the polarization and depolarization of neurons that have the molecular target of this drug. Possibly those populations of neurons that possess the NMDA receptors are located mainly in areas of the secondary associative cortex, which would explain the observed results. Another hypothesis that should be considered is that Ketamine compromised the process of communication and transmission of information between several areas of the cortex (the communication process from $cortex' \longrightarrow cortex''$). Once in the secondary associative cortex, the primary source of inputs comes from other areas of the cortex; a decrease in the capacity of transmission of stimuli between distinct cortical regions would have a more accentuated effect on secondary associative areas.

\subsubsection*{Assumptions: Promoted Areas}

An increase in the presence of functional clusters and, posteriorly, an intensification of functional connectivity were prominently observed in regions respective to the primary sensory and motor cortices.

The delimited patterns of the clusters, and the expressive registers of those processes most of the time, suggest that they have a non-spontaneous nature. Taking into account the way they were presented and the principles of organization of the primary cortex, it is possible to suggest that the main source of inputs of those regions, \textbf{the thalamus, was involved with the observed activities}.

Experiments of positron emission tomography analyzing the effects of the administration of Ketamine in human patients revealed an increase in the metabolic activity of the cingulate cortex, thalamus, and putamen \citep{laangsjo2003effects}, whereby the most prominent alterations being a considerable increase in the thalamus activity   \citep{laangsjo2004effects}.

The influence of thalamic contributions on cortical activities, mainly those in the primary sensory cortex, is constantly debated in neuroscience. However, there are experimental reports that demonstrate how thalamic stimuli are able to influence activities that occur in those regions of the cortex. For example, McLean using a calcium imaging technique mapped the activities of large populations of cells in the fourth layer of the primary visual cortex in mice \citep{maclean2005internal}.
From the characterization of Spatio-temporal patterns of spontaneous and thalamic-triggered activities, it was observed that the Spatio-temporal patterns of both types of activities demonstrated to be statistically indistinguishable. Those results demonstrate that thalamic stimuli act as a trigger able to evoke patterns of activities that are intrinsic from cortical circuits \citep{maclean2005internal}. Maclean has concluded that \textbf{successions of thalamic} \textbf{inputs can ``awaken'' the cortex}, confirming a similar hypothesis made previously by Sherman \citep{sherman2001tonic}.

The characteristics of the primary sensory and motor cortex, the experimental observations and reports of McLean that thalamic inputs can awaken the cortex, the experimental reports of L{\aa}ngsj{\"o} that the administration of Ketamine in humans induces an increase in the activities of the thalamus support the hypothesis that \textbf{the thalamus was involved with the activities observed in the primary cortex} during Ketamine-Medetomidine general anesthesia.

The analysis of the results, mainly the observations during the transition from the anesthetized to the awake state (the process of recovering), in which were possible to observe patterns sometimes characteristic of the awake state and times characteristic of the anesthetized state, seems to suggest that thalamic inputs activating areas of the primary cortex constitute a compensatory mechanism used by the brain to re-establish the activities in the secondary associative cortex. Indicating that, besides processing visual and motor information, the primary cortex may also have the function of giving support and helping to sustain the balance of activities occurring in the secondary associative cortex.
 
 \vspace{2.9\baselineskip}

\section{Conclusions}

The present study provided experimental evidence that large-scale functional neural networks do possess a modular organization character. Furthermore, the fact that it was broadly present on all frequency bands analyzed and in both awake and anesthetized conditions highlights the presence of functionally integrated modular neural processes as a crucial aspect of functional brain activities, which supports the consensus that this organization structure may account for a series of properties necessary for the brain's functioning.  
The supposed view that specialized areas of the cortex operate preferably in an isolated manner is not consistent with the experimental results of this study. Instead, verifying the structure of functional clusters involving large cortical areas supports the view that, essentially, several cortical areas operate in conjunction forming an integrative network, and not in an isolated manner.
It was observed that functional brain activities are dynamic. Furthermore, the community structure over short time intervals presented variations in participating regions, size, and number of clusters, revealing that the working brain is highly complex and dynamic.
The experimental results of this study demonstrated that during the Ketamine-Medetomidine anesthetic induction, considerable alterations in the community structure of functional brain networks occurred; the most affected areas were the secondary associative cortex of the frontal and parietal regions. The effects of the anesthetics on those regions may have significantly contributed to the loss of consciousness experienced by the animal model in this study.

It was verified that during general anesthesia, the macaque brain had shifted to a highly specific, complex, and dynamic state, characterized by a massive presence of functional clusters at the primary sensory and motor cortices, and a reduction of these integrated and segregated neural processes on the secondary associative cortex. Those empirical results do not support suppositions that regard general anesthesia with a ``whole-brain shut-down''; once during the experiment, the brain seemed to be performing much more than homeostasis.
The manner in which the functional clusters manifested in the primary sensory and motor cortices, the principles of organization of the cerebral cortex, the experimental reports, and statements in the literature support the supposition that the thalamus was involved in those activities observed. In addition, the careful observation of the structure and dynamics of functional modules during the experiment suggested that the primary cortex may also perform a function to sustain and support the dynamic balance of functional activities in other areas of the cerebral cortex.

\bibliography{dissertacao}


\end{multicols}

\end{document}